\documentclass{aa}
\usepackage{graphicx}
\begin{document}
   \title{Effects of correlated turbulent velocity fields on the
           formation of \\  maser lines }


   \author{R. B\"oger\inst{1}\and W.H. Kegel\inst{2,3} \and M. Hegmann\inst{3}}

   \offprints{R. B\"oger}
   \institute{Hamburger Sternwarte, Universit\"at  Hamburg,
               Gojenbergsweg 112, D-21029 Hamburg\\
              \email{rboeger@hs.uni-hamburg.de}
         \and
            Institut f\"ur theoretische Physik der Universit\"at
            Frankfurt am Main, Robert-Mayer Strasse 8-10,\\
            D-60054 Frankfurt am Main\\
             \email{kegel@astro.uni-frankfurt.de}
         \and
            Zentrum f\"ur Astronomie und Astrophysik der Technischen Universit\"at
            Berlin,\\ Sekr. PN 8-1, Hardenbergstrasse 36,
            D-10623 Berlin\\
             \email{[hegmann,kegel]@astro.physik.tu-berlin.de}
             }            

   \date{Received 19 February 2003 /Accepted 20 May 2003}

   \abstract{
The microturbulent approximation of turbulent motions is widely used
in radiative transfer calculations. Mainly motivated by its 
simple computational application it is probably in many cases an oversimplified
treatment of the dynamical processes involved.
This aspect is in particular important in the analysis of maser lines, 
since the strong amplification of radiation 
leads to a sensitive dependence of the radiation field on the overall velocity structure.
To demonstrate the influence of large scale motions on the formation of maser 
lines we present 
a simple stochastic model which takes velocity correlations into 
account. For a quantitative analysis of correlation effects, we generate 
in a Monte Carlo simulation individual realizations of a turbulent 
velocity field along a line of sight. Depending on the 
size of the velocity correlation length we find huge deviations between 
the resulting random profiles in respect of line shape, intensity and position 
of single spectral components. Finally, we simulate the emission of extended
maser sources. A qualitative comparison with observed masers associated
with star forming regions shows that our model can reproduce the 
observed general spectral characteristics.  We also investigate shortly,
how the spectra are effected when a systematic velocity field (simulating
expansion) is superposed on the fluctuations. Our results  convincingly demonstrate 
that hydrodynamical motions are of great importance for 
the understanding of cosmic masers.

             }

   \maketitle
%

\section{Introduction}             

Maser lines of various molecular species are observed in star forming
regions and in the envelopes of evolved late type stars. In both cases it is
well known that turbulent motions play an important role in the gas
hydrodynamics. Their spectra show large velocity differences (up to tens of 
km/s) and are variable on observable time scales. VLBI-observations reveal
that maser regions often consist of several isolated sources with different 
radial velocities (see, e.g., Reid et al. \cite{Reid}).
Due to the Doppler effect, the radiation field depends on the local velocity 
structure. This coupling to the velocity field does sensitively
affect the  formation of maser lines since in the unsaturated regime the
radiation is exponentially amplified. To the extent that stochastic 
hydrodynamical velocities are involved,
the correlation length, i.e., the length over which the hydrodynamical velocity
changes substantially, plays an important role. In the usual 
microturbulent analysis large scale motions are neglected. As a consequence, 
each spectral component will  generally be  attributed 
to a separate source and could in principle be formed under different physical conditions.
The aim of the  present investigation is to emphasize the necessity to consider
hydrodynamical motions in more detail.
We will show how correlated velocity fields affect the line forming process and that
omitting this important aspect can lead to erroneous interpretations of 
observed maser spectra.

Traving and collaborators (Gail et al. \cite{Gail}; 
Traving \cite{Traving}) developed a theory allowing to account for
correlated turbulent motions\footnote{ Here and in the following we use the term turbulence
in the spectroscopic sense (which is wider
than in hydrodynamics) referring to any stochastic motion leading to line
broadening. The terms micro-, meso-, and macroturbulence refer to
the way in which the turbulent motions affect the line broadening which
depends on the ratio of the mean free path of a photon to the correlation
length of the velocity field.} in the transfer of line radiation.
In their formalism, the hydrodynamic velocity $v$ along a line of sight is described by a
Markovian process defined by a Gaussian distribution function and an
exponential correlation function. Due to the stochastic modeling of the
velocity field, the intensity also has to be considered as a random
variable. Traving showed that in accord with these assumptions and together
with the usual equation of transfer, a Fokker-Planck equation can be derived,
which describes the probability $W(I_{\nu },v,s)$ of finding at
point $s$, the intensity $I_{\nu }$ and the velocity $v$. Of particular
interest is the expectation value of the intensity $\left\langle I_{\nu
}\right\rangle $ since it reflects as a time or spatial average the mean
physical properties of the system. Gail et al. (\cite{Gail2}) examined the model of
an unsaturated maser applying the theory described above. 

While this approach gives a first qualitative insight into the effects introduced
by a turbulent velocity field, the question arises to which extent the calculated
expectation value of the intensity may be directly compared with observed spectra.
The answer depends in particular on two issues. The first is the question of how many
statistically independent lines of sight contribute to a given observation. This
obviously is a function of the spatial resolution. The second point is the question
of how many spectra corresponding to individual lines of sight have to be averaged
in order to obtain the expectation value with a certain accuracy. As will be shown
later, in the case of unsaturated masers this number is very large. In order to investigate
these questions 
we set up a Monte Carlo (MC)
simulation scheme to generate individual realizations of the random velocity field. The
calculated line profiles reflect the stochastic nature of the turbulent motion
along a given line of sight and allow a quantitative analysis of correlation
effects.

\section{Model assumptions and the Monte Carlo method}

In the case of an unsaturated maser the occupation numbers of the maser levels 
are essentially not affected by the intensity in the maser line. 
Consequently, the intensity variation along
any line of sight depends just on the local physical conditions.
In order to focus on the effect of correlated
velocity fields, we assume a plane parallel slab with a homogeneous density
and temperature distribution. Additionally, we take the velocity field to be
quasi-static, presuming that the dynamical time scales are large in comparison 
with the duration of individual observations.
The variation of the turbulent velocity along each line of sight is assumed to follow
a  Markov process and the
projected velocity $v(s)$ along a line of sight is treated as a random
function of $s$. The one-point distribution function of $v$ is considered to
be Gaussian with the rms turbulent velocity $\sigma_\mathrm{t}$ 
\begin{equation}
W\left(  v\right)  =\frac{1}{\sqrt{2\pi}\sigma_\mathrm{t}}\cdot\:\exp{\left[
-\frac{v^{2}}{2\sigma_\mathrm{t}^{2}}\right]} \,.
\end{equation}
The conditional probability of finding at point $s+\Delta s$ the velocity $v+\Delta v$
is 
\begin{eqnarray}
\lefteqn{P\left(  v+\Delta v,s+\Delta s\right|  \left.  v,s\right)  = {}}
  \nonumber\\
& &{}\mbox{\hspace{.9cm}} \frac
{1}{\sqrt{2\pi\sigma_\mathrm{t}^{2}\left(  1-f^{2}\right)  }}\:\exp\left\{
-\frac{\left[  \Delta v+v(1-f)\right]  ^{2}}{2\sigma_\mathrm{t}^{2}\left(
1-f^{2}\right)  }\right\}  \label{condprob}
\end{eqnarray}
with an exponential type correlation function 
\begin{equation}
f\left(  \Delta s\right)  =\:\exp\left[  -\frac{\left|  \Delta s\right|
}{l}\right] \,. 
\end{equation}
   \begin{figure}
   \centering
   \resizebox{\hsize}{!}{\includegraphics[clip]{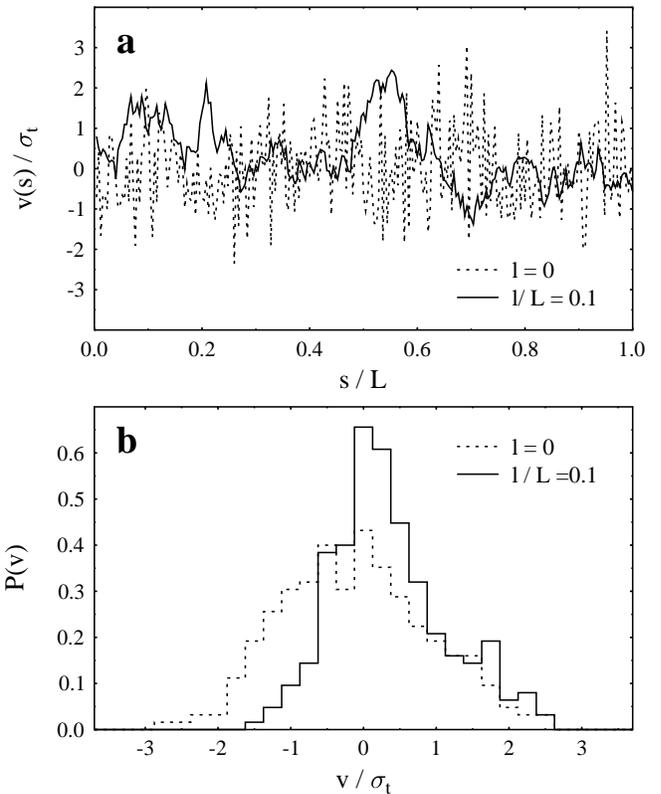}}
      \caption{(a) Individual realizations of the one-dimensional velocity
                    field, and (b) the corresponding distribution function
                    $P\left(v\right)$ for the microturbulent ($l=0$, dashed histogram)
                    and a mesoturbulent  ($l/L=0.1$, solid histogram) model
              }
         \label{fig1}
   \end{figure}
The correlation length $l$ defines the length scale of the stochastic velocity 
variation.  From Eq.~(\ref{condprob}) it follows that the expectation value of $\Delta v$ is
\begin{equation}
\left\langle \Delta v\right\rangle =-v\left[  1-\:\exp\left(
-\frac{\left|  \Delta s\right|  }{l}\right)  \right]
\end{equation}
and the variance is given by
\begin{equation}
\mathrm{Var}\left(  \Delta v\right)  =\sigma_\mathrm{t}^{2}\left[  1-\:\exp\left(
-\frac{2\left|  \Delta s\right|  }{l}\right)  \right]  \,.
\label{VardeltaV}%
\end{equation}

 The assumption that the variation of the velocity field along the line of sight
may be described by a Markov process (implying an exponential correlation function)
is the simplest way to introduce correlation effects into the theory of radiative
transfer (see e.g. Gail et al. 1974, 1980). This assumption implies that only the
one-point and the two-point distribution functions are important. All other correlation functions
require the knowledge of higher order multi-point distribution functions. For later
comparison with the work of others, we note that
the (spatial) power spectrum of the velocity field is the Fourier transform of the 
correlation function. This means that (3) implies a one-dimensional power spectrum

\begin{equation}
P_1(k) \sim  \frac{{l}}{1 + {l}^2k^2}
\end{equation}
which for ${l}^2k^2 \gg 1$ is well approximated by a power law, $P(k) \sim k^{-2}$. If we
consider (3) as correlation function in three dimensions, i.e. if we consider $s$ as a vector,
we find for the three-dimensional power spectrum
 \begin{equation}
 P_3(k) \sim  \frac{{l}}{(1 + {l}^2k^2)^2}
\end{equation}

The
choice of the correlation function (3) makes it relatively easy to calculate individual realizations
of the velocity field (see below), as well as the expectation value of the line profile using
the generalized radiative transfer equation (Gail et al. 1974, 1975).

In accord with Eq.~(\ref{condprob}) the distribution of the random projected velocities can 
be generated in a Monte Carlo procedure (Levshakov et al. \cite{Levshakov2}):
\begin{equation}
v\left(  s+\Delta s\right)  =\xi\sqrt{\sigma_\mathrm{t}^{2}\left(  1-f^{2}\right)
}+fv\left(  s\right) \,,
\end{equation}
where $\xi$ is a normally distributed random number with $\left<\xi\right>=0$ and 
$\mathrm{Var}(\xi)=1$. The velocity at the edge of the cloud is given by $v=\sigma_\mathrm{t}\xi(s=0)$.

The line forming process depends sensitively on the ratio $\sigma_\mathrm{t} / v_\mathrm{th}$ 
(see, e.g., Levshakov \& Kegel \cite{Levshakov}).
In the case  $\sigma_\mathrm{t} / v_\mathrm{th}<1$, the line broadening is thermally 
dominated and the turbulent velocity correlation is insignificant. Contrary, if 
$\sigma_\mathrm{t} / v_\mathrm{th} >1$ correlation effects become pronounced. An inappropriately chosen step 
size can smear out the correlated structure of a random velocity field. This is 
avoided by satisfying the inequality
\begin{equation}
\mathrm{Var}\left(  \Delta v\right)  <v_\mathrm{th}^2 \,.
\end{equation}
From (\ref{VardeltaV}) it follows that for $\sigma_\mathrm{t}>v_\mathrm{th}$ this is equivalent to
\begin{equation}
\frac{\left|\Delta s\right|}{l}\le\ln\left[  1-\left(v_\mathrm{th}/ \sigma_\mathrm{t}
\right)^{2}\right]^{-1/2} \,.
\end{equation}
To illustrate the influence of a finite correlation length, Fig.~\ref{fig1} shows one
realization of a completely uncorrelated  (microturbulent) velocity field ($l=0$, dashed line)
and one realization  of a mesoturbulent velocity field ($l>0$, solid line).
The lower panel represents the corresponding projected velocity distributions 
$P\left( v \right)$. As 
expected, the velocity in the limiting case is normally distributed, whereas the 
correlated velocity distribution deviates significantly from a Gaussian. This is 
due to the fact that for a finite ratio $l/L$ the statistical base represents an 
incomplete statistical sample. Averaging over many realizations (either temporal 
or spatial) would yield an complete  ensemble with a unique Gaussian 
distribution function.
   \begin{figure}
   \centering
   \resizebox{\hsize}{!}{\includegraphics[clip]{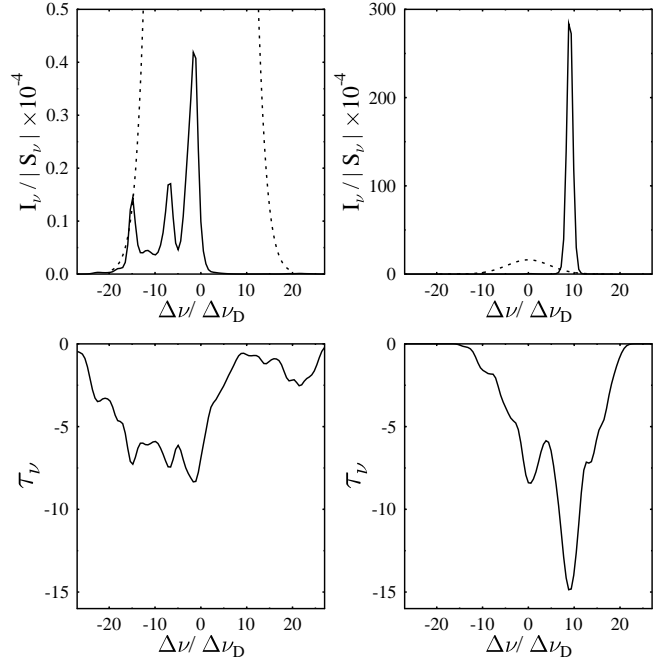}}
      \caption{Upper panels: Line profiles for two individual realizations of the velocity field
                (solid curve) 
                for the parameter set $l/L=0.1,  \sigma_\mathrm{t} / v_\mathrm{th} = 10$, 
                and $\tau_0=-110$. For comparison also the spectrum of the intensity expectation value
                is shown (dashed curve). Lower panels: The optical depth of the random
                spectra 
              }
         \label{fig2}
   \end{figure}
   \begin{figure}
   \centering
   \resizebox{\hsize}{!}{\includegraphics[clip]{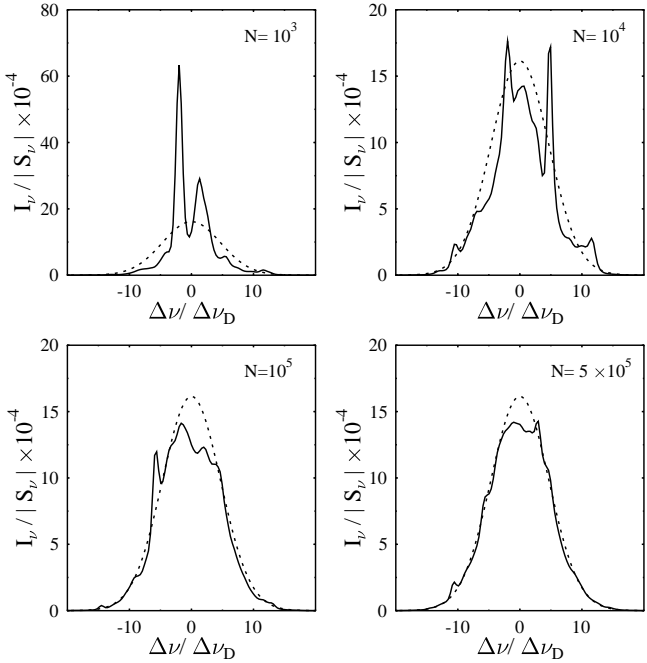}}
      \caption{Convergence of the ensemble average of N random profiles (solid line)
                to its expectation value (dashed line). The parameter configuration
                is the same as in Fig. \ref{fig2} 
              }
         \label{fig3}
   \end{figure}
For simplicity, we neglect the dependence of the occupation numbers of the involved
maser levels on the radiation in the maser line. If furthermore the source 
function $S_\nu$ is assumed to be constant over the whole region, the equation 
of radiative transfer can be solved analytically. Neglecting the background radiation yields
\begin{equation}
I_{\nu}\left(  s=0\right)  =S_{\nu}\left\{  1-\:\exp\left[  -\tau_{\nu
}\left(  L\right)  \right]  \right\} \,, \label{S-T-Gl3}
\end{equation}
 with the optical depth 
\begin{equation}
\tau_{\nu}\left(  s\right)  =\int\limits_{0}^{s}\kappa_{\nu}ds' \,,
\end{equation}
and the monochromatic absorption coefficient
\begin{equation}
\kappa_{\nu}=\kappa_{0}\Phi\left(  \Delta\nu,v\right)  
\label{kappa-ny} \,,
\end{equation}
$\Phi\left(  \Delta\nu,v\right)$ being the local profile function.
It should be noted that for population inversion the source function and the 
absorption coefficient are both negative. If one considers thermal Doppler 
broadening as the dominant effect, the local profile function can be written as
\begin{equation}
\Phi\left(  \Delta\nu,v\right)
=\frac{1}{\sqrt{\pi}\Delta\nu_\mathrm{D}}\cdot\:\exp\left\{  -\left[
\frac{\left(  \nu-\nu_{0}\right)  }{\Delta\nu_\mathrm{D}}-\frac{v}{v_\mathrm{th}}\right]
^{2}\right\} \,,
\end{equation}
where $\Delta\nu_\mathrm{D}$ is the thermal Doppler width and $v_\mathrm{th}$ the thermal 
velocity. Equally, the projected random velocity distribution along a line of sight $P\left(v\right)$
can be convolved with the thermal profile function to derive the optical depth
\begin{equation}
\tau_\nu\left(L\right)= \kappa_0 L \hspace{-.15cm}
\int\limits_{-\infty}^{\infty}
P\left(v\right)\Phi\left(  \Delta\nu,v\right)dv\,.\label{convolut}
\end{equation}

\section{Results}
Our model calculations mainly depend  on three parameters: the optical
thickness of the slab $\tau_0$, the turbulent velocity $\sigma _\mathrm{t}$ and the
correlation length $l$. Due to the random nature of the velocity field, the optical
thickness can not be defined uniquely. Therefore, it is characterized by the
optical depth in the line center for vanishing turbulent motions, 
$\sigma_\mathrm{t}=0$, which is then given by $\kappa_0 L$.

Figure~\ref{fig2} demonstrates the influence of velocity correlations on the line forming
process. The upper panels show the line profiles for two individual realizations
of the velocity field along a given line of sight calculated with the same statistical parameters (solid curves). 
For comparison, also the
expectation values of the intensity are shown (dashed curves). The lower panels give the corresponding
optical depths $\tau_\nu$. In our simple model (constant density and constant pumping
efficiency), the latter reflect directly the complex structure
of the underlying velocity field (see Eq.~[\ref{convolut}]). Due to the exponential 
amplification  the intensity profiles appear to have a much simpler structure than 
the $\tau_\nu$-profiles. This is in particular true if a lower intensity
cut-off is introduced.
Therefore do high intensity spectra often consist  of a single line only,
which may exhibit a pronounced line shift.    

As is obvious from Fig.~\ref{fig2} (see also Table~\ref{table1}), the line profiles calculated for
individual realizations of the velocity field can differ substantially from each other
and from the expectation value.
The degree to which an individual realization can deviate from the expectation value can
statistically be described by the convergence behavior of the ensemble average. We 
find that in general the mean value approaches the expectation value very slowly. For 
the chosen parameter configuration, Fig.~\ref{fig3} shows a sequence of averages
over $N$ realizations. Even the superposition of $5\times 10^5$ individual spectra  
is not sufficient to match the spectrum of the intensity expectation value. 
For a smaller correlation length or
larger turbulent motions the convergence is improved but still a substantial number 
of profiles is required. This finding is contrary to the case of pure absorption. In the 
context of quasar absorption lines Levshakov et al. (\cite{Levshakov2}) showed that 
the average of about $100$ realizations reproduces
the expectation value of the intensity within a one percent error margin.
   \begin{figure*}
   \centering
   \resizebox{\hsize}{!}{\includegraphics[clip]{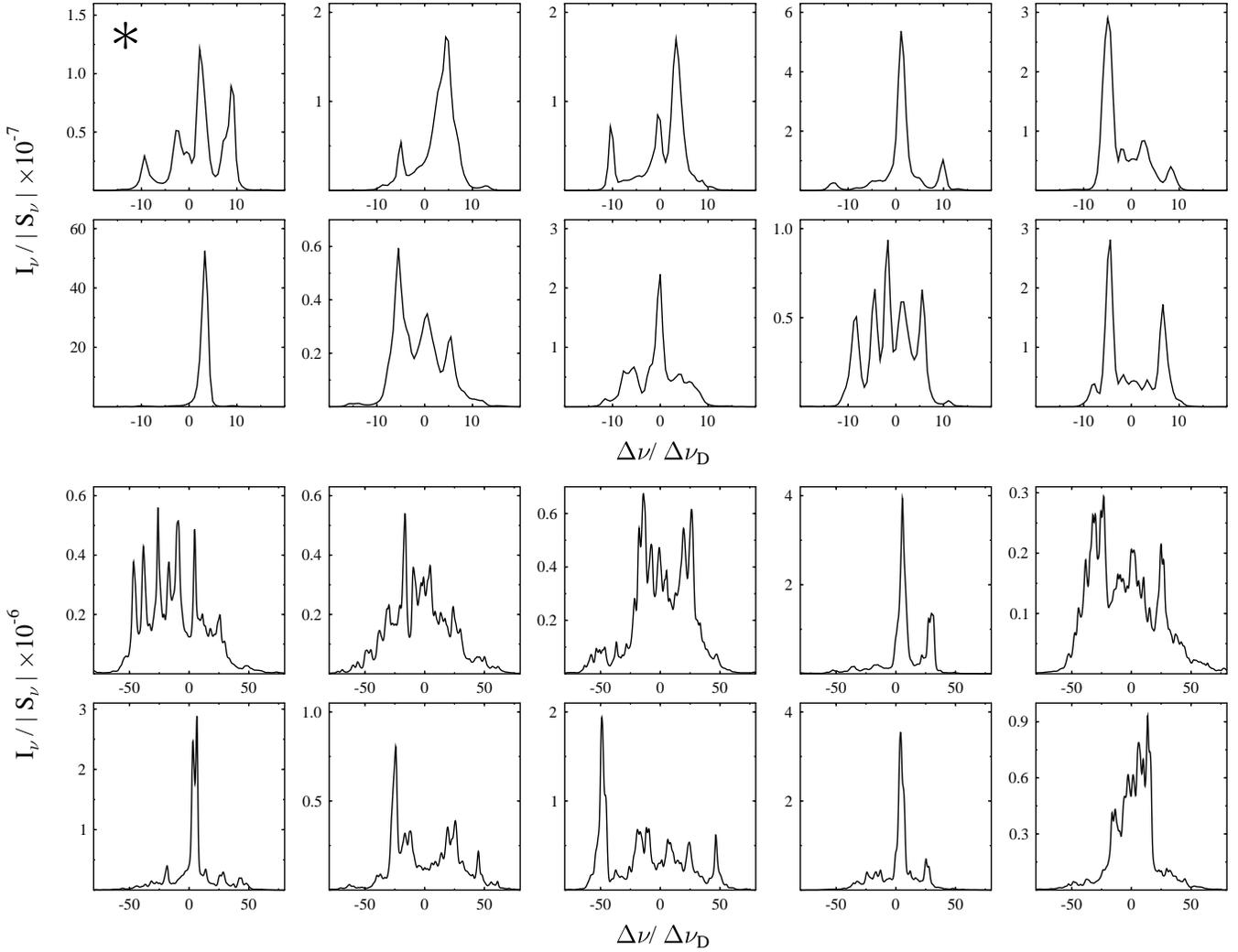}}
      \caption{Simulated spectra of an extended maser region.
                Each spectrum is composed of one hundred single realizations. In the  upper
                ten panels we show the model calculations with a parameter set
                of $l/L=0.1$, $\sigma_\mathrm{t} / v_\mathrm{th} = 10$, $\tau_0=-110$ and 
                in the lower ten panels the parameter set is
                $l/L=0.1$,  $\sigma_\mathrm{t} / v_\mathrm{th} = 50$, $\tau_0=-400$.
                 The spectrum marked by an asterisk is used in Fig.~\ref{fig8}
              }
         \label{fig4}
   \end{figure*}
The radiation of an extended source reflects the average influence of the dynamical motions on the
radiation field. From the above finding it can  be concluded that in the case of an
 unsaturated maser the expectation value of the intensity will in general not be in good
agreement with the measurements if a finite correlation length is involved. Even if the 
problem is solved  self-consistently, where, due to saturation effects, 
the deviations from the mean intensity will be less pronounced, it can be expected that 
the mean intensity will not match the observations.   Despite of this conclusion, 
the method by Traving is a valuable tool to examine the general influence of the stochastic 
parameters which govern the dynamical motions. It allows a very fast calculation of
expectation values and distribution functions (Levshakov et al. \cite{Levshakov2}). 
In this respect, the method is preferable
to a MC simulation which requires the generation of a very large number of realizations and 
is thus  costly in computer time.
Starting from the same model assumptions, Gail et al. (\cite{Gail2}) examined the mean properties of the 
radiation field. Their most important finding is that the mean intensity increases with  increasing
velocity correlation. This corresponds to our result that with a larger correlation
length the probability of having large peak intensities grows (see Table 
\ref{table1}). 
   \begin{table}
      \caption[]{For various correlation length the probability  that the maximum intensity
                  of an individual random spectrum exceeds a certain value is given. For 
                  all cases the peak intensities are compared to the intensity expectation 
                  value in the line center as calculated from the parameter configuration
                  given in Fig.~\ref{fig2}}
         \label{table1}
           \begin{center}
           \begin{tabular}{rrrr}
            \hline
            \noalign{\smallskip}
            Greater than & $L/l=10$ & $L/l=25$ & $L/l=50$ \\
            \noalign{\smallskip}
            \hline
            \noalign{\smallskip}
           $0.1\cdot \left\langle I_{\Delta\nu=0}\right\rangle $ & $69.2 \%$ & $  37.3 \%$ & $  15.7 \%$  \\
           $1\cdot\left\langle I_{\Delta\nu=0}\right\rangle$      &$ 28.0 \%$      & $3.5 \%$     & $0.13 \%$\\
           $10\cdot \left\langle I_{\Delta\nu=0}\right\rangle$      & $6.5 \%$      & $0.06 \%$      & $0.0 \%$ \\
           $100\cdot \left\langle I_{\Delta\nu=0}\right\rangle$ &      $0.9 \%$ &      $0.0 \%$    & $0.0 \%$  \\
            \noalign{\smallskip}
            \hline
            \end{tabular}
            \end{center}
   \end{table}  
   \begin{figure}
   \centering
   \resizebox{\hsize}{!}{\includegraphics[clip]{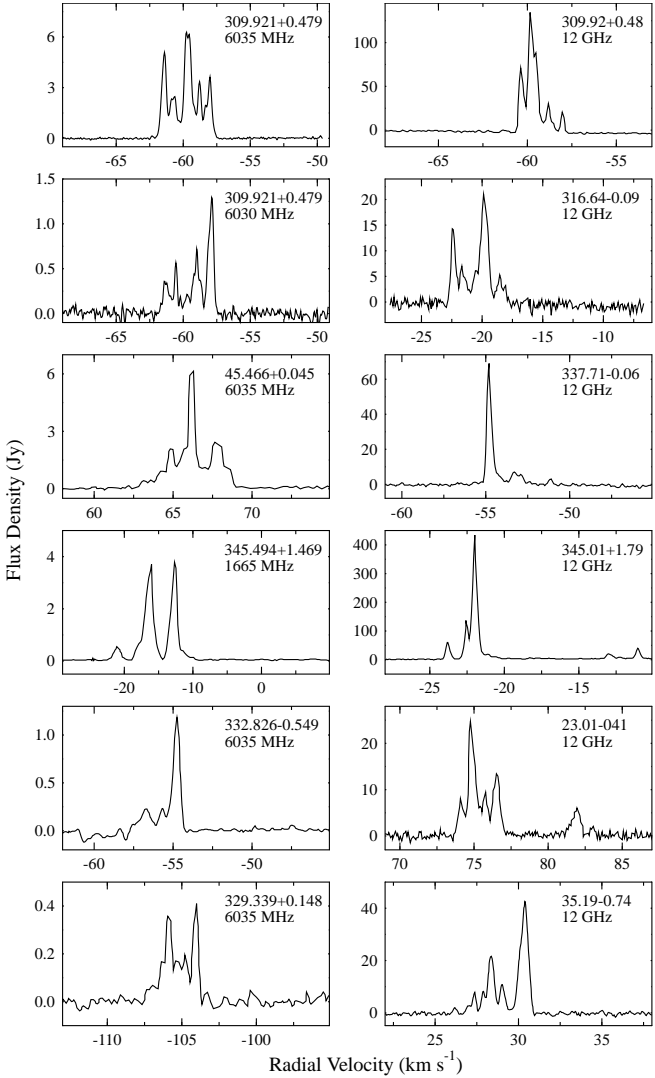}}
      \caption{A sample of selected $OH$ (left column) and $CH_3OH$ (right column)
                maser spectra observed by J. Caswell.
                All sources are associated with star forming  regions
                or ultra compact $\rm\ion{H}{ii}$ regions. Methanol masers
                are in general  negligibly polarized which makes them particularly
                suitable for a comparison with our synthetic spectra. Except 
                of the two spectra of the source 309.921+0.479 which are 
                lefthanded polarized  the polarization of all other OH sources
                is unknown              
                }
         \label{fig5}
   \end{figure}  
   \begin{figure}
   \centering
   \resizebox{\hsize}{!}{\includegraphics[clip]{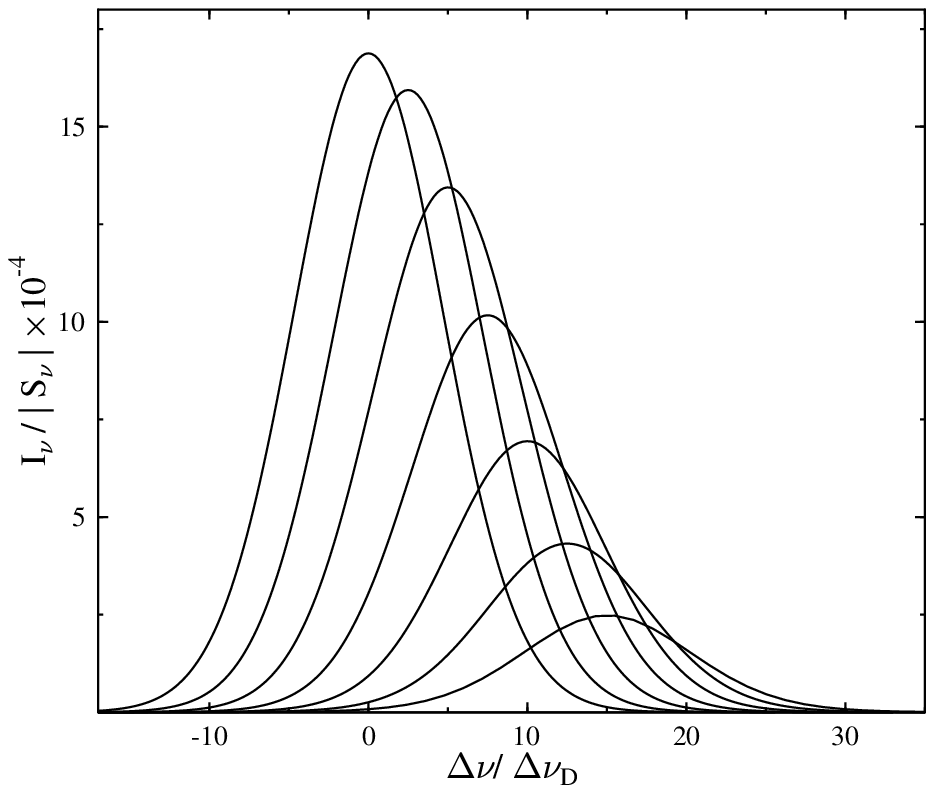}}
      \caption{Expectation values of the intensity calculated for an expanding
                plane parallel slap with a constant velocity gradient of
                $dv_\mathrm{sys}/ds = 0, 5, 10, 15, 20, 25, 30\, v_\mathrm{th}/L$
                (in decreasing order). All other parameters are the same as in Fig.~\ref{fig2}
              }
      \label{fig6}
   \end{figure}
In the present paper we want to focus on a different aspect of the problem
for which the MC technique allows a more quantitative comparison of observations and model 
calculations.
In the framework of our simple model, it is reasonable to approximate the radiation of
an extended maser region by the superposition of several spectra corresponding to statistically
 independent lines of sight, i.e. to different realizations of the stochastic velocity field.
This condition is met sufficiently if the distance between two lines 
of sight is larger than one correlation length. A number of $L/l \times L/l$ lines 
of sight will then be adequate if the transverse extent of the region is
comparable to its depth. Figure~\ref{fig4} shows some composite spectra of  $100$ realizations,
with $l/L=0.1$ and $\sigma_\mathrm{t}=10\times v_\mathrm{th}$ and $\sigma_\mathrm{t}=50\times v_\mathrm{th}$
respectively. 
It should be emphasized that these spectra
are not selected by predefined criteria but are randomly chosen. It is conspicuous that they
exhibit a distinct structure which is far from the spectrum of the intensity expectation value.
They show 
pronounced maxima spreading partly over a frequency range of more than $2\times\sigma_\mathrm{t}$. 
In the standard analysis they would be attributed to several independent sources at different radial 
velocities.
As mentioned before, the spiky structure of the spectra is caused by the correlations in the
velocity field. It should be noted, however, that - due to the exponential 
amplification - the intensity distribution does not reflect directly the velocity 
distribution (see Fig.~\ref{fig2} in which the optical depth 
is proportional to the projected velocity distribution).

Figure~\ref{fig5} shows a selected sample of observed $OH$ and $CH_3OH$ maser spectra obtained 
by J. Caswell (private communication). A detailed description of most sources can be found
in Caswell (\cite{Caswell1}, \cite{Caswell2}, \cite{Caswell3}) together with additional 
maser spectra. Our simple model does
clearly not allow a detailed analysis of the observations.
However, it is remarkable that it can reproduce the characteristic structure of the observed
line profiles. Also the velocity scale of our model spectra is in  agreement with the 
observations, if for example, a typical  cloud temperature of 25 K is assumed. 
In this case the thermal velocity of the $OH$ and $CH_3OH$ molecules is 
$v_\mathrm{th}(OH) = 0.156 \,\mathrm{km s}^{-1}$ and $v_\mathrm{th}(CH_3OH) = 0.114 \,\mathrm{km s}^{-1}$,
respectively.
Due to the higher velocity dispersion, 
the spectra in the lower panels of Fig.~\ref{fig4} show a richer structure and span
a larger velocity range than the spectra in the top panels. Of course,  also 
the amplification of radiation is affected by larger velocity differences along a line of sight.
Therefore, we had to chose for both models very different 
opacity parameters to achieve the typical spiky structure of maser profiles.

In this context, one may also ask  how an additional systematic velocity field 
influences the line formation. For this purpose we considered an expanding 
plane parallel slab with a constant systematic velocity gradient superposed on the stochastic
velocity field. The general effect of 
this modification of our model can be viewed in Fig.~\ref{fig6} where the expectation 
value of the intensity for various values of the expansion velocity is shown (solid lines). 
As expected,  this additional velocity component leads on average to a 
weakening of the radiated emission and to a shift of the velocity distribution function.
Additionally, a slight broadening of the profile of the intensity expectation value is seen. Its
width gives directly an estimation of the velocity range over which individual random 
spectra will be distributed.
   \begin{figure}
   \centering
   \resizebox{\hsize}{!}{\includegraphics[clip]{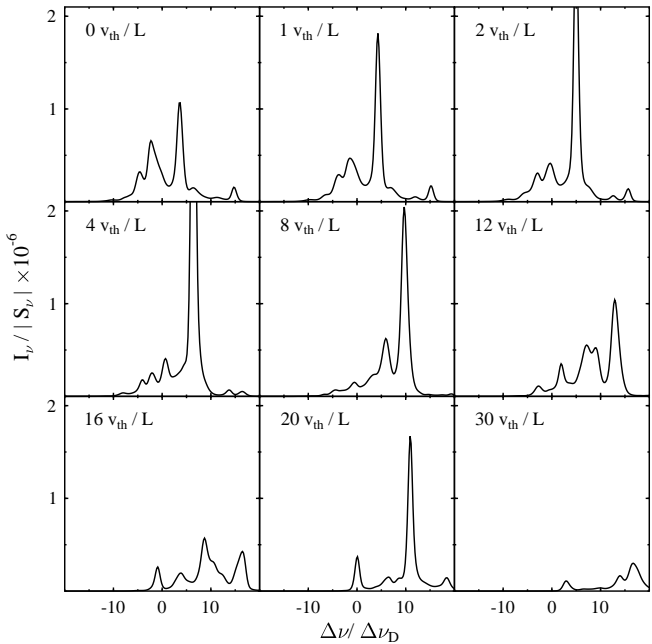}}
      \caption{As in Fig.~\ref{fig4} the spectra model the emission of an extended maser region with
                the parameter configuration:
                $l/L=0.1$, $\sigma_\mathrm{t} / v_\mathrm{th} = 10$, $\tau_0=-110$.
                For all calculations the same turbulent velocity field was used.
                Additionally we consider along the line of sights a systematic velocity
                component with a constant velocity gradient $dv_\mathrm{sys}/ds$. 
                The values used in each calculation are shown in the upper left corner
                of the panels 
              }
               \label{fig7}
   \end{figure}
The shape of individual line profiles is very sensitive to changes of the velocity 
field. In Fig.~\ref{fig7} we model again the radiation of a maser region by a superposition of 
$L/l \times L/l$ lines of sight. The turbulent velocity field is in all cases the same and only
the strength of the linear expansion differs. From this example it is obvious 
that even small outflow velocities can significantly change the appearance of the resulting 
spectra. The tendency that with increasing systematic motions the emission
is reduced is also clearly seen. But the gradients of the systematic and the turbulent
velocity field can  also partly compensate each other, allowing a strong amplification of 
radiation along particular lines of sight. This behavior is seen in the sequence 
$dv_\mathrm{sys}/ds = 0, 1, 2, 4~v_\mathrm{th}/L$ and  in the case 
$dv_\mathrm{sys}/ds = 20~v_\mathrm{th}/L$.

\begin{figure}
   \centering
   \resizebox{\hsize}{!}{\includegraphics[clip]{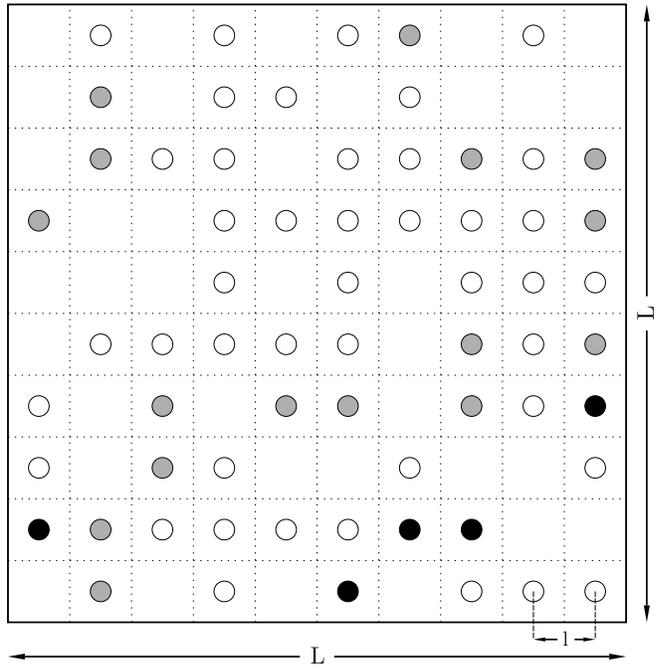}}
      \caption{Spatial intensity distribution of the spectrum in Fig.\ref{fig4}
                marked with an asterisk. Black circles indicate a maximum intensity
                of a single realization greater than ten times the expectation
                value in the line center.  Grey and white
                circles show maximum intensities in the range between 
                $10\left< I_{\nu=0}\right>\geq I_\mathrm{\nu ,max} \geq
              \left<I_{\nu=0}\right>$ and $\left<I_{\nu=0}\right>
              \geq I_\mathrm{\nu ,max} \geq 0.1\left<I_{\nu=0}\right>$ respectively.
                The intensity in the blank boxes is below one tenth of the
                mean intensity in the line center  
              }
         \label{fig8}
   \end{figure}

It is also instructive to simulate the spatial intensity distribution of composite spectra.
As described before the
emerging radiation from our model cloud is approximated by the superposition
of $L/l \times L/l$ independent lines
of sight corresponding to individual realizations of the velocity field. 
The general characteristics of the spatial intensity map of the cloud's surface can be simulated 
by distributing the maximum intensity found in  each contributing spectrum randomly on a square grid
of $L/l \times L/l$ cells. In Fig.~\ref{fig8} we consider as an example the spectrum marked with an 
asterisk in Fig.~\ref{fig4}. The black circles indicate 
maximum intensities greater than ten times the expectation value in the line center.
Their number corresponds to the number of distinct spectral components. The grey and white
circles show maximum intensities in the range between 
$10\left< I_{\nu=0}\right>\geq I_\mathrm{\nu ,max} \geq \left<I_{\nu=0}\right>$ and  
$\left<I_{\nu=0}\right>\geq I_\mathrm{\nu ,max} \geq 0.1\left<I_{\nu=0}\right>$ respectively.
The maximum intensity in the blank boxes is below one tenth of the expectation value of
the intensity in the line center. Due to the underlying stochastic velocity field,
large spatial intensity
variations are seen. 
Similar to observed high resolution intensity maps the maser region consists
of a few strong sources at 
different radial velocities (see, e.g., Reid et al. \cite{Reid}), while
the radiation from most of the region is inconspicuous. 
\section{Conclusion}

In the previous sections we have investigated the influence of a
turbulent velocity field with finite correlation length on the appearance
of a cosmic maser source. For this, we used a very simple stochastic model
in which the only spatially varying quantity is the hydrodynamical velocity,
while all other parameters, in particular the density and the pump efficiency
are held constant. With this simple model, we could to a surprisingly high degree
reproduce the observed characteristics 
of cosmic masers, in particular those related to star forming regions. Our results
are a strong indication that the details of the hydrodynamic velocity field
inside a maser region are of great importance for the interpretation of the 
observed properties of the source. Of course, reality is considerably more
complex than our simple model. It is very likely that in star forming regions
not only the velocity is a fluctuating quantity, but also the density   and other
physical parameters determining the pump efficiency show strong spatial variations.
However, our results convincingly show that the observation of a maser spot, by itself does
not necessarily imply that the physical conditions at the location of the spot
are peculiar. The spot may as well be the result of the details of the velocity
distribution along the line of sight.

In view of the simplicity of our model, we have made no attempt to compare results 
in more detail with observational results of particular maser regions. Such attempts, however,
have been performed by other authors investigating effects of turbulent motions on 
maser spectra. The approach closest to ours is that of Sobolev et al. (1998).
They consider also a slab which is homogeneous except for the velocity field.
The latter is characterized by its statistical properties. From these, individual
realizations of the velocity field are constructed which then are used to calculate the
optical depth and the emitted spectrum. Their approach differs from ours in the assumptions about 
the velocity field and in the numerical procedure. For the velocity field they assume a 
more general power spectrum
\begin{equation}
P_3(k) \sim \frac {k^2}{(k_\mathrm{min}^2 + k^2)^\alpha} ,
\end{equation}
where $k_\mathrm{min}$ has been introduced to avoid divergence for $k \rightarrow 0$. Physically,
$k_\mathrm{min}$ is identified with the inverse of the linear size of the emitting volume.
(In our case this identification would correspond to the assumption ${l}/L = 1$.) To construct
individual
realizations of the velocity field random sampling is done in k-space (for details see
Dubinski et al., 1995) and the velocity field is then obtained by a Fourier transformation.
Sobolev et al. investigate in particular models with $\alpha$ close to $17/6$. The case $\alpha = 17/6$
corresponds to a Kolmogorov spectrum, which is the common approximation for turbulence in
an incompressible fluid. The general result obtained by Sobolev et al. is similar to ours:
The spiky structure of maser spectra and the concentration of the emitted radiation in narrow
bright spots, may be caused solely by the velocity field. - They then compare their results
with detailed observations of the methanol masers in OMC-1. They find evidence that the
power spectrum of the turbulence in OMC-1 is somewhat steeper than an actual
Kolmogorov spectrum and conclude that $\alpha \ge 3$. (With $\alpha = 3$ the spectrum
(16) approaches (7) in the limit $k \rightarrow \infty$) - We note in passing that for 
this more general velocity field
there exists no generalized radiative transfer equation as was derived 
by Traving and associates ( Gail et al. 1974, Traving 1975) for the special correlation function (3). I.e., 
if one is interested in the expectation value of the
spectrum, one actually has to calculate the average of a sufficiently large number
of spectra for individual realizations. In the case of maser emission this number 
may be very large, as shown in Fig.~\ref{fig3}.

A very different approach is followed by Gwinn (1994a,b). He tries to solve the
inverse problem, i.e. he attempts to derive the characteristics of the velocity 
field directly from the very detailed observations available for the $H_2O$ maser region in W49N.
The observations are interpreted by a model in which a strong stellar wind strikes 
ambient material and the arising shocks provide the excitation energy for the masers. 
More than 250 maser features were identified. Gwinn measures the distribution of the
individual maser spots in coordinate and velocity space. From the derived power law
correlation functions he concludes that the velocity field in the maser region is
turbulent. This interpretation of the observations makes the (implicit) assumption
that the velocities of the individual maser spots reflect directly the velocity 
field in the maser region. This assumption is appropriate if one considers the
maser spots as physical entities like clumps. If one considers models like ours
or that of Sobolev et al. (1998), in which the maser spots are solely caused by
the correlations in the velocity field, the situation is more complex (see Fig.~\ref{fig2}). - A
similar statistical analysis of high quality observational data for the $H_2O$ masers
in the star forming regions Sgr B2(M), W49N, W51(MAIN), W51N, and W3(OH) has been performed 
by Strelnitski et al. (2002). They find that the two-dimensional distribution of
maser spots shows a fractal structure and that the two-point velocity structure functions
can be approximated by power laws with exponents close to Kolmogorov's values. From these 
findings they conclude that the velocity field in the these maser regions is highly turbulent.              

\begin{acknowledgements} 
We thank Dr. James Caswell for providing the observed maser
spectra shown in Fig.~\ref{fig5} and for  many interesting discussions. We also acknowledge
constructive comments of the referee Dr. C. R. Gwinn, as well as discussions with
Dr. B. Deiss and C. Hengel on the different types of correlation functions and 
power spectra.\\
This research has been partly supported by the BMBF/DLR
under Grant No. 50 OR 0203.
\end{acknowledgements}


\begin{thebibliography}{}
  \bibitem[1993]{Caswell1} Caswell, J. L. 1993 MNRAS, 260, 425
  \bibitem[1998]{Caswell2} Caswell, J. L. 1998 MNRAS, 297, 215
  \bibitem[2001]{Caswell3} Caswell, J. L. 2001 MNRAS, 326, 805
  \bibitem[1995] {Dubinski} Dubinski, J., Narayan, R., \& Phillips, T. G. 
                  1995, APJ 448, 226
  \bibitem[1974]{Gail} Gail, H.-P., Hundt, E.,Kegel, W. H., Schmid-Burgk, J., \&
                  Traving, G. 1974, A\&A, 32, 65
  \bibitem[1975]{Gail2} Gail, H.-P., Kegel, W. H., \&  Sedlmayr, E.
                  1975 A\&A, 42, 81
  \bibitem[1980]{Gail3} Gail, H.-P., Sedlmayr, E., \& Traving, G. 1980 JQRST, 23, 267
  \bibitem[1994a]{Gwinn1} Gwinn, C. R. 1994a APJ 429, 241
  \bibitem[1994b]{Gwinn2} Gwinn, C. R. 1994b APJ 429, 253
  \bibitem[1994]{Levshakov}Levshakov, S. A., \& Kegel, W. H. 
                  1994, MNRAS, 271, 161
  \bibitem[1997]{Levshakov2}Levshakov, S. A., Kegel, W. H., \& 
                        Mazets, I. E. 1997, MNRAS, 288, 802
  \bibitem[1980]{Reid}Reid, M. J., Haschick, A. D., Burke, B. F. et al.
                         1980, APJ, 239, 89
  \bibitem[1998]{Sob} Sobolev, A. M., Wallin, B. K., \& Watson, W. D. 1998, APJ 498, 763
  \bibitem[2002]{Strel} Strelnitski, V., Alexander, J., Gezari, S. et al. 2002, APJ 581, 1180
  \bibitem[1975]{Traving} Traving, G. 1975, in Baschek, B., Kegel, W. H., \&
                  Traving, G., eds. Problems in Stellar Atmospheres and Envelopes
                  (Springer, Berlin), 326                        




\end{thebibliography}
\end{document}